# Solar Activity and Space Weather

**Nat Gopalswamy[1], Pertti Mäkelä[1,2], Seiji Yashiro[1,2], Sachiko Akiyama[1,2], and Hong Xie[1,2]**

[1.] NASA Goddard Space Flight Center, Greenbelt, MD 20771, USA
[2.] The Catholic University of America, Washington DC 20064, USA

nat.gopalswamy@nasa.gov

**Abstract**. After providing an overview of solar activity as measured by the sunspot number (SSN) and space weather events during solar cycles (SCs) 21-24, we focus on the weak solar activity in SC 24. The weak solar activity reduces the number of energetic eruptions from the Sun and hence the number of space weather events. The speeds of coronal mass ejections (CMEs), interplanetary (IP) shocks, and the background solar wind all declined in SC 24. One of the main heliospheric consequences of weak solar activity is the reduced total (magnetic + gas) pressure, magnetic field strength, and Alfvén speed. There are three groups of phenomena that decline to different degrees in SC 24 relative to the corresponding ones in SC 23: (i) those that decline more than SSN does, (ii) those that decline like SSN, and (iii) those that decline less than SSN does. The decrease in the number of severe space weather events such as high-energy solar energetic particle (SEP) events and intense geomagnetic storms is deeper than the decline in SSN. The reduction in the number of severe space weather events can be explained by the backreaction of the weak heliosphere on CMEs. CMEs expand anomalously and hence their magnetic content is diluted resulting in weaker geomagnetic storms. The reduction in the number of intense geomagnetic storms caused by corotating interaction regions is also drastic. The diminished heliospheric magnetic field in SC 24 reduces the efficiency of particle acceleration, resulting in fewer high-energy SEP events. The numbers of IP type II radio bursts, IP socks, and high-intensity energetic storm particle events closely follow the number of fast and wide CMEs (and approximately SSN) because all these phenomena are closely related to CME-driven shocks. The number of halo CMEs in SC 24 declines less than SSN does, mainly due to the weak heliospheric state. Phenomena such as IP CMEs and magnetic clouds related to frontside halos also do not decline significantly. The mild space weather is likely to continue in SC 25, whose strength has been predicted to be not too different from that of SC 24.

## 1. Introduction

Observational manifestations of solar activity are the appearance and dispersal of closed and open magnetic field regions on the Sun, represented by sunspot regions and coronal holes, respectively. Thus, solar activity characterizes magnetic variability of the Sun and is often measured by indices such as the sunspot number (SSN) and the radio flux at 10.7 cm wavelength (F10.7). Disturbances emanating from closed magnetic regions are flares and coronal mass ejections (CMEs), while high-speed streams (HSS) originate from coronal holes. CMEs and HSS interact with the ambient solar wind that lead to the formation of shock sheaths ahead of CMEs (if fast enough) and stream interaction regions (SIRs) at the leading edge of HSS. When SIRs continue for more than a solar rotation, they are called corotating interaction regions (CIRs). These disturbances cause various space weather events such as sudden ionspheric disturbance by flare photons, solar energetic particle (SEP) events, and geomagnetic storms (see recent reviews [1-4]).

This article is concerned with the mild space weather in solar cycle (SC) 24, characterized by the reduced number and frequency of space weather events as compared to SC 23. In particular, we consider large SEP events (>10 MeV proton intensity ≥10 pfu) and intense geomagnetic storms (Dst ≤ –100 nT). We compare these numbers with SSN and the number of energetic CMEs. CMEs originate from sunspot and non-spot regions, so the relation between SSN and CME rate is somewhat complicated [5]. On the other hand, energetic CMEs that cause space weather events generally originate from sunspot regions (manifestation of the toroidal field of the Sun) because only these regions can store and release large amounts of magnetic energy [6]. Weaker CMEs seem to be closely related to the global (poloidal) magnetic field of the Sun [7-8]. Reduced global field strength results in reduced heliospheric pressure that affect the size and magnetic content of all CMEs, including those from sunspot regions. Thus both the toroidal and poloidal components of the solar magnetic field affect CMEs. While CMEs occur in high numbers during solar maximum, low-latitude coronal holes occur more frequently in the declining phase of the SC (see e.g., [9]) and hence geomagnetic storms caused by CIRs dominate in the decliling phase. In this article, we examine the properties of coronal holes underlying intense geomagnetic storms. Finally, we address the question of whether mild space weather will continue in SC 25 by estimating the cycle strength using microwave polar brightening.

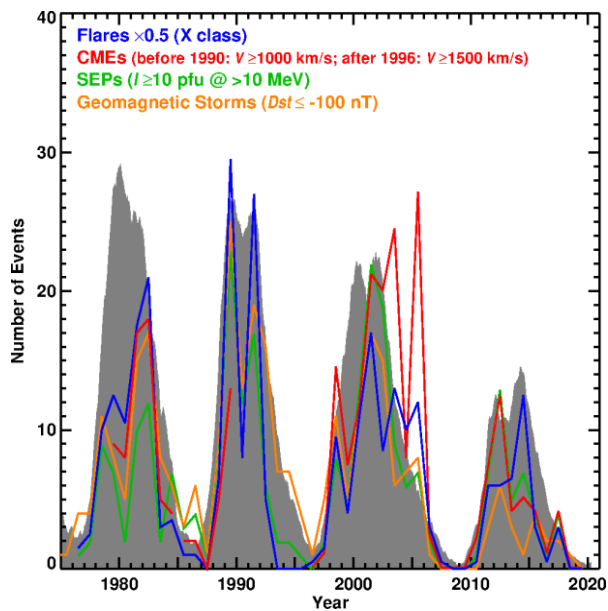

**Figure 1.** Annual numbers of solar events relevant to space weather (X-class solar flares and fast CMEs) and space weather events (SEP events, geomagnetic storms) superposed on V2.0 SSN (grey, arbitrary scale) during SCs 21-24. Large SEP events with >10 MeV proton intensity ≥10 pfu, and major geomagnetic storms (Dst ≤ – 100 nT), are included. The number of GOES X-class flares is divided by 2 to fit the scale. CME data are from the Solwind coronagraph (1979-1985), Solar Maximum Mission Coronagraph/Polarimeter (1985-1989), and SOHO/LASCO (1996-2019).

## 2. Solar Activity Cycle and Space Weather Events

Figure 1 shows the solar cycle variation of solar events (X-class flares and fast CMEs) and related space weather events (large SEP events and major geomagneic storms) along with SSN during SCs 21-24. All cycles show double peaks in the maximum phase, although the second peak is barely discernible in SC 21. One expects a relation between SSN and energetic CMEs because only sunspot regions possess large amounts of free energy needed to power such CMEs. Even though there is overall similarity of various numbers to SSN, marked discordance can be seen between CME and flare numbers. In detail, there are many differences. Of the 4 cycles shown, all numbers vary according to SSN only during both peaks of SC 22. In SC 21, the number of eruptions (flares and CMEs) and space weather events peak before and after the SSN peaks. In SC 23, all numbers peak along with the second SSN peak, but well before the first peak. In SC 24, the CME and flare numbers have discordant behavior during both peaks. During the prolonged minimum between cycles 23 and 24, all numbers are consistently zero. There are several reasons for the discordance. While more free energy can be stored in sunspot regions, energetic CMEs can also originate from outside active regions causing large SEP events [10-12]. Fast CMEs can originate anywhere on the disk (even backsided), but SEP events require magnetic connectivity to the particle detectors. Similarly, only fast CMEs originating close the disk center directly impact Earth and cause geomagnetic storms, provided the underlying flux rope has a southward component of the magnetic field either in the rope or in the leading sheath. Some X-class flares can be confined and hence are not associated with a CME. Many SEP events and geomagnetic storms are produced by CMEs associated with M and C-class flares.

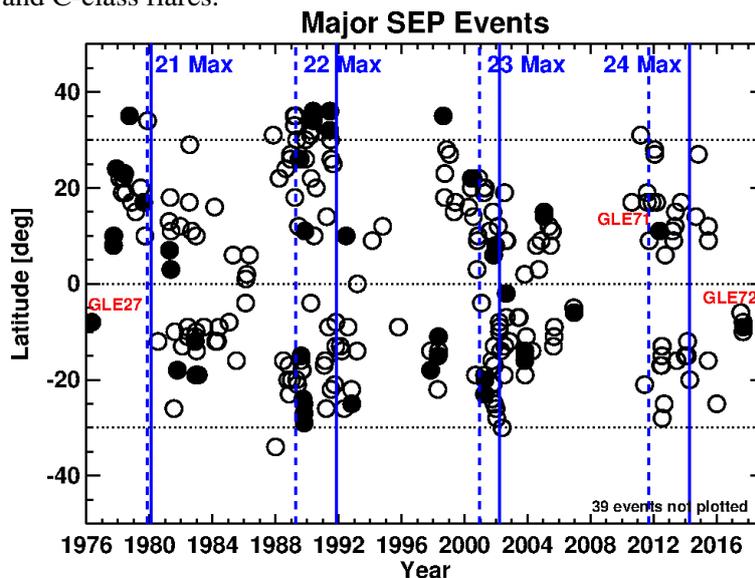

Figure 2. Locations of large on-disk SEP events that occurred during the recent four SCs 21-24 (1976 to 2019). The dominant maxima of these cycles are marked by the vertical solid blue line. The secondary peak is denoted by the dashed vertical line. In cycles 22-14, the dominant peak corresponds to excess sunspots in the southern hemisphere. In SC 21, the rising phase up to the maximum had more sunspots in the north, briefly switching to south dominance in 1980. However, there were smaller SSN peaks in 1981 and 1982 with north and south dominance, respectively. SEP clusters can be found in 1981 in the north and in 1982 in the south. SEP events with ground level enhancement (GLE) are distinguished by the filled circles. The clustering of the data points around the solar maxima is evident (updated from [6]).

The discordance between SSN and fast CMEs is more pronounced between the two SSN peaks (2012 and 2014) of SC 24 [13]. There were 6 major geomagnetic storms in 2012, compared to just one in 2014. Similarly, there were 15 large SEP events in 2012 compared to just 7 in 2014. The number of halo CMEs is substantially higher with 84 in 2012 and 63 in 2014. The number of interplanetary type II bursts were similar during the two peaks (19 in 2012; 16 in 2014), however, the corresponding average CME speeds are very different: 1543 in 2012 vs. 1201 km/s in 2014. The speed difference explains the higher number of large SEP events during the 2012 peak. While the number of halo CMEs originating from the disk center (likely to cause geomagnetic storms) are nearly the same in the two peaks (17 in 2012; 14 in 2014), their average speeds are different: 975 km/s in 2012 vs. 753 km/s in 2014. The speed difference explains the low geoeffectiveness in the 2014 peak because CME speed is one of the key factor deciding the strength of a geomagnetic storm. In contrast to all these events, the number of X-class flares is only 7 in 2012 compared to 16 in 2014. Even if we exclude the three CMEless X-class flares, the number of X-class flares is a factor 2 higher in 2014 than in 2012. It appears that the CME kinetic energy revealed by interplanetary type II bursts is a better indicator of space weather events than the X-class flares.

Figure 2 shows the distribution of solar-source latitudes of CMEs that resulted in large SEP events during SCs 21 to 24. It is clear that the SEP sources occur in clusters in both hemispheres. The clusters tend to occur around a sunspot peak, either dominant (solid vertical line) or secondary (dashed blue line). The dominant peak in each of the four cycles corresponds to the southern hemisphere (see http://sidc.be/silso). Recall from Fig. 1 that the SEP source clusters in SC 21 were on either side of the time of the peak SSN. In Figure 2 we see three clusters in SC 21: 1978 (northern hemisphere), 1981 (northern hemisphere), and 1982 (southern hemisphere). The three clusters are associated with three smaller SSN peaks, but not the main peak, which is a sharp spike. In SC 22, the first SSN peak has more spots from the northern hemisphere, but SEP clusters are present in both hemispheres. The dominant second peak has more spots in the southern hemisphere, consistent with the SEP cluster there. In SC 23, SEP events occur in large numbers throughout the cycle. There are two large clusters one in the north (secondary peak) and the other in the south (dominant peak). In SC 24, the secondary peak in 2012 is mainly contributed by spots in the northern hemisphere and is associated with an SEP cluster. The second dominant SSN peak in 2014 has SEP events both from the northern and southern hemispheres. It is thus clear that SEP events (including GLE events) are more abundant around peaks of SSN, but occasionally occur away from the peaks due to energetic active regions.

An important observational fact evident in Figure 2 is that there are only 2 GLE events in SC 24 in contrast to about a dozen events in other cycles. This is the drastic reduction in high-energy SEP events highlighted in [14]. The two SC-24 GLEs are on 2012 May 17 (northern hemisphere) and 2017 September 10 (southern hemisphere). In addition, there was a sub GLE event on 2014 January 6 from the southern hemisphere [15] that was backsided (39 backsided SEP events are not shown in Fig. 2). SEP events will be further discussed below.

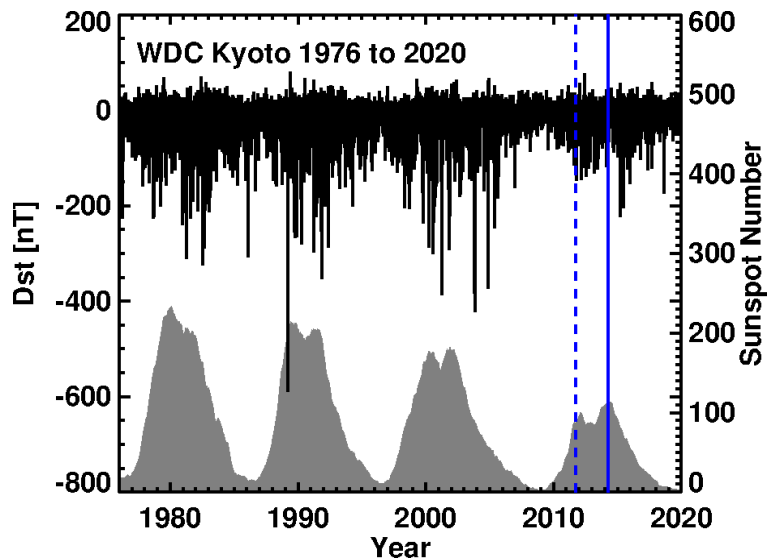

**Figure 3.** Dst index and SSN as a function of time during SCs 21-24. The first and second peaks of SC 24 are marked by the vertical blue lines. Each downward spike corresponds to a geomagnetic storm. Spikes extending to and below –100 nT are considered intense storms. The spikes extending to Dst >0 are sudden commencements or sudden impulses. The data are from the World Data Center, Kyoto.

Figure 3 shows the long-term variation of the Dst index during SCs 21-24 showing the level of geomagnetic activity during these cycles. The largest storm in the space era occurred on 1989 March 13 in SC 22. As in the case of SEP events, SC 23 witnessed geomagnetic storms throughout the cycle. The SC 24 stands out in being very low in geomagnetic activity [6,16]. As noted in Fig.1, the CMEs from around the 2012 peak of SC 24 were more geoeffective than those around the 2014 peak. The top three storms of the cycle occurred in the declining phase: two from sunspot regions (–222 nT on 2015 March 17 and –204 nT on 2015 June 21) and one from a filament region (–175 nT on 2018 August 26).

## 3. CME Occurrence Rate and Sunspot Number

We now take a closer look at cycles 23 and 24 in understanding the relation between CME daily rate and SSN. The different behaviors of CME rate and average speed between SCs 23 and 24 are shown in Fig. 4. Here we consider CMEs with width ≥30⁰ to exclude poor events. It is clear that the daily CME rate in SC 24 is highly fluctuating, but overall higher in SC 24. On the other hand, the average speed is substantially lower in SC 24. Even within SC 24, the speed is higher in the first SSN peak (2012) than in the second (2014), confirming what we concluded above. Figure 4 also shows that both the daily rate and the average speed of CMEs have the clear appearance of solar cycle variation. However, the relationship between SSN and CME rate has very different behavior in SC 23 and 24. Lamy et al. [17] reported steeper slopes in SC 24 between CME rates and other activity indices such as F10.7. Figure 4 shows scatter plots between SSN and CME rate in SC 23 and 24. While the two parameters are highly correlated, the slopes of the regression lines are very different: the SC 24 slope is higher by a factor of 2 [18], indicating the higher number of CMEs in SC 24.

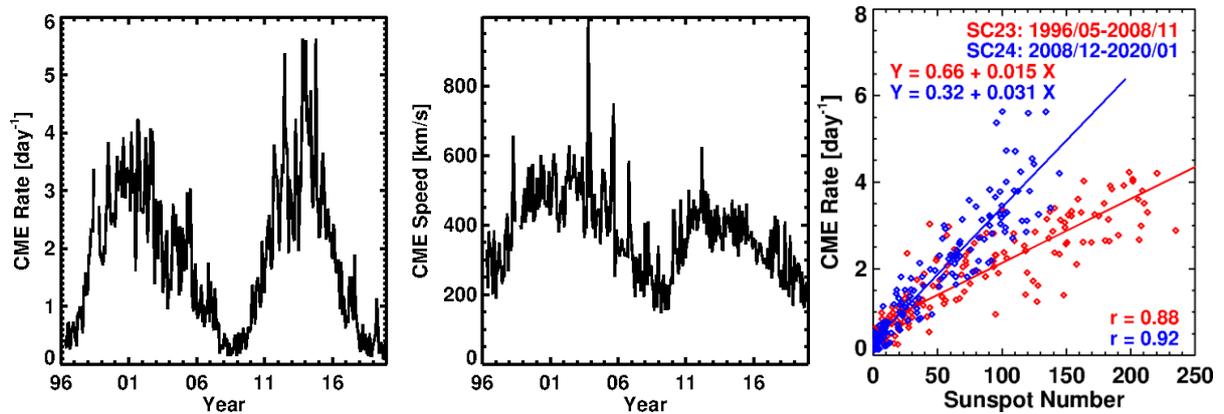

**Figure 4**. The daily occurrence rate (left) and speed (middle) of SOHO/LASCO CMEs with width ≥30⁰. Both quantities are averaged over Carrington rotation periods. The plotted speed is computed as an average of all CMEs occurring in a Carrington rotation period. The large spikes are due to super-active regions that produced many CMEs in quick succession. (right) Scatter plots between SSN and CME daily occurrence rate in SC 23 (red) and SC 24 (blue). The SSN range in the two cycles indicate a weaker SC 24.

The changed relation between SSN and CME rate in SC 24 triggered a debate on the reason behind the change. Luhmann et al. [19] explained enhanced CME rate of SC 24 in terms of the weak polar fields indicating less constraint on closed magnetic field, so more eruption occurred, and the CMEs were able to escape. The weakened state of the heliosphere [14] allows smaller CMEs to appear bigger and counted. This reason is related to the Luhmann et al. [19] suggestion because the heliospheric total pressure drops when the global magnetic field strength drops. Some thought the enhanced rate is possibly an artifact related to the cadence change in SOHO/LASCO observations that occurred in 2010 August [20-21]. However, Petrie [7] examined CMEs reported by manual and automatic catalogs of CMEs and showed that the enhanced CME rate in SC 24 actually started with the polarity reversal of SC 23 in 2004 followed by the establishment of weaker polar fields. Recently, Michalek et al. [8] showed that the enhancement is due to a population of weak CMEs that occurred since 2004 in addition to the regular CMEs that follow the pattern of SSN. They also demonstrated that the enhanced CME rate is due to a significant decrease of total (magnetic and plasma) heliospheric pressure as well as the changed magnetic pattern of solar corona.

The different heliospheric states in the two cycles is demonstrated in Fig. 5. Most parameters except the solar wind density differed significantly between the two cycles. The cycle averages roughly correspond to the maximum phase. The total pressure, magnetic field strength, and the Alfven speed monotonically decrease from 2004, reach a minimum in 2009 and then slowly rises to reach the cycle-average level in the maximum phase of SC 24. This decline is consistent with the conclusions made in [7-8] regarding the overabundance of CMEs in SC 24. The behavior of the Alfven speed is similar to that of the magnetic field because the density is approximately constant. While slightly slower (by ~7%), the SC 24 solar wind is cooler by ~23%.

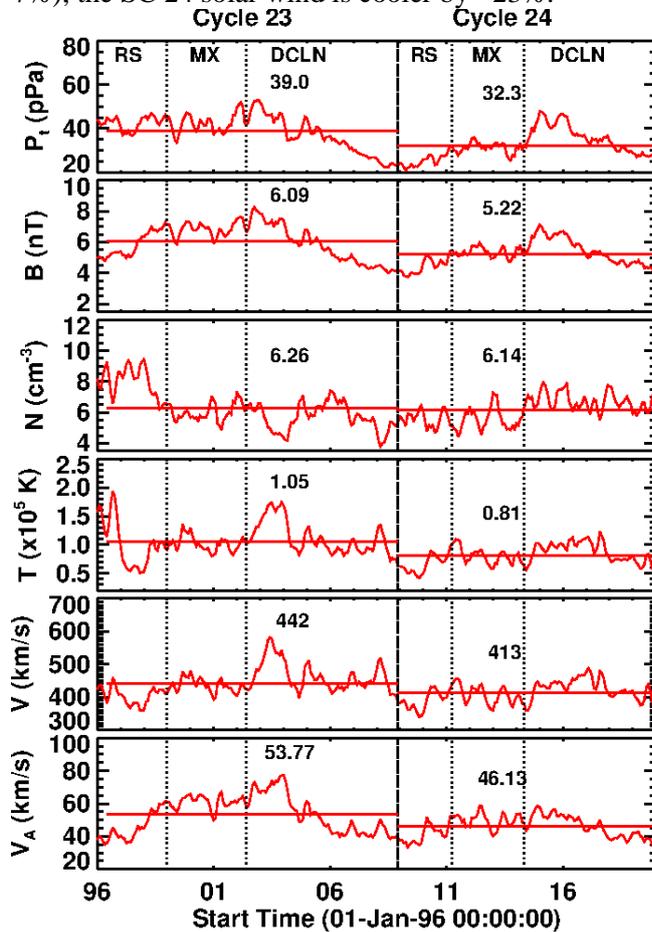

**Figure 5.** Solar wind parameters in SCs 23 and 24: Total pressure ($P_t$), magnetic field strength (B), proton density (N), proton temperature (T), solar wind background speed (V), and Alfven speed ($V_A$). The vertical dotted lines delineate the rise (RS), maximum (MX), and declining (DCLN) phases of each cycle. The vertical dashed line separates the two cycles. The cycle averages of various parameters are denoted by the horizontal bars with the average values noted on the plots.

## 4. Impact of Weak Solar Activity in Cycle 24 on Space Weather

In the middle of SC 24, it was recognized that space weather is mild in this cycle [6,14,22-24] due to the reduced heliospheric magnetic field and total pressure (see Fig. 5). The weak heliospheric state manifests in a number of ways: (i) for a given CME speed, the CME width in the coronagraph FOV is larger in SC 24 [14,18], (ii) the number of halo CMEs did not decrease substantially in SC 24 and halos originated at larger central meridian distances [25-26]; (iii) limb CMEs attained halo status sooner and at lower speeds in SC 24 than in SC 23 [18,27]; (iv) CME expansion speed measured in the corona is ~48% higher in SC 24 [28]; (v) SC 24 CMEs attained constant width at a larger heliocentric distance [23].

The effect of the weak solar cycle on the number and severity of space weather events has been shown compounded by the backreaction of the heliosphere on CMEs and their shocks [14,18,23,27]. The altered properties of CMEs resulted space weather that is milder than what is expected from the reduced activity alone. The same applies to space weather caused by CIRs [4,29,30]. Instead of looking at the intense geomagnetic storms, if one starts with the magnetic clouds (MCs) and examines their space weather impact, the reduction in geoeffectiveness (as measured by Dst) is again clear [23,30]. The geoeffectiveness of SC 24 MCs relative to the ones in SC 23 declined by 49%; the geoeffectiveness of sheaths ahead of MCs declined by 59% [23]. Yermolaev et al. [30] showed the reduced geoeffectiveness to be similar in non-cloud ejecta and sheath. In addition, these authors compared the geoeffectiveness of various solar wind structures that occurred in SCs 21-24 and found drastic reduction of geoeffectiveness in SC 24 compared to that in other cycles. The general pattern is that the reduction in geoeffectiveness is far deeper than the reduction in solar activity. To further illustrate this, we show the time variation of the number of halo CMEs and fast and wide (FW) CMEs in Fig. 6 and 7, respectively. Figure 7 also compares the time variation of the number of FW CMEs (Fig. 7a) with those of geomagnetic storms (Fig. 7b), SEP events (Fig. 7c), and type II bursts in the decameter-hectometric (DH) wavelength range.

Table 1. Comparison of various parameters between solar cycles 23 and 24

| Property | Cycle 23 | Cycle 24 | Ratio |
|---|---|---|---|
| Cycle-averaged SSN | 81 | 49 | 0.61 |
| # Halo CMEs | 396 | 324 | 0.82 |
| # CMEs V≥900 km/s & W≥60° | 485 | 253 | 0.52 |
| Average limb CME speed (km/s)[a] | 689 | 568 | 0.82 |
| # ≥M1.0 flares | 1568 | 798 | 0.51 |
| # ≥C1.0 flares | 14730 | 8676 | 0.59 |
| # ICMEs | 307 | 208 | 0.68 |
| # MCs | 114 | 86 | 0.75 |
| # IP shocks (fast forward)[b] | 272 | 143 | 0.52 |
| Average IP shock speed (km/s)[c] | 482 | 374 | 0.79 |
| Fast Mach number[c] | 2.14 | 2.04 | 0.95 |
| # Dst ≤ - 100 nT storms (all) | 86 | 22 | 0.26 |
| # Dst ≤ - 100 nT storms (ICME) | 77 | 20 | 0.26 |
| # Dst ≤ - 100 nT storms (CIR) | 12 | 3 | 0.25 |
| # ≥10 pfu GOES SEP events | 102 | 46 | 0.45 |
| # SEP-ESP events (GOES) | 17 | 9 | 0.53 |
| # GLE events | 16 | 2 | 0.13 |
| # DH Type II bursts | 339 | 181 | 0.53 |
| SW Total Pressure (pPa) | 39.0 | 32.3 | 0.83 |
| SW Magnetic field strength B (nT) | 6.09 | 5.22 | 0.86 |
| SW Bulk speed V km/s | 442 | 413 | 0.93 |
| SW Proton density N (cm$^{-3}$) | 6.26 | 6.14 | 0.98 |
| SW Proton temperature T ($10^5$ K) | 1.05 | 0.81 | 0.77 |
| SW Alfven Speed $V_A$ (km/s) | 53.77 | 46.13 | 0.86 |

[a] Associated with limb flares of size ≥C3.0; [b] From Wind data base; [c] Excludes 19 events with Fast Mach number <1 and 2 events with shock speed ≤0.

Table 1 summarizes the reduction in various space weather events and the associated solar events in cycles 23 and 24. For reference, we have given the cycle-averaged SSN. The last column in Table 1 gives the ratio of a parameter in SC 24 and 23, so subtraction of the ratio from 1 gives the drop in the value of the parameter in SC 24. For example, the ratio of SSN is 0.61, which indicates that SSN dropped by 39% in SC 24 (from 81 in SC 23 to 49). The solar wind (SW) parameters are from Fig. 5. The number of flares with soft X-ray intensity ≥C1.0 declined by 41%, similar to the decline in SSN. However, at higher intensities the decline is more: the number of ≥M1.0 flares dropped by 49%. Comparing partial cycles 23 and 24, Alberti et al. [31] found the number of ≥M2.0 flares by declined by 40%. CMEs associated with flares of size ≥C3.0 are used for avoiding projection effects in determining the near-Sun speed of CMEs. High-energy SEP events and intense geomagnetic storms show the sharpest decline in SC 24. These and other space weather events are further discussed below.

*4.1. Halo CMEs*
Figure 6a shows the number of halo CMEs in each cycle summed over Carrington rotation periods. There were 396 full halos in SC 23, which lasted for 151 months from May 1996 to November 2008.

There was a 4-month SOHO data gap. If we assume that halo CMEs in the gap interval occurred at the same rate as the rest of the time, the total number of cycle-23 halos is expected to be 407. In SC 24, there were 323 halos, indicating a drop of only 21%, which is half of the drop in SSN. Since the SSN averaged over cycles 23 and 24 are 81 and 49, respectively, we see that SC 24 has more halo CMEs per SSN. The higher abundance of halo CMEs in SC 24 seems to be due to the anomalous expansion of CMEs that makes it easy for a CME to attain halo status. Interestingly, Fig. 6b,c show that the average speed of halo CMEs has decreased from 1625 km/s in SC 23 to 1162 km/s in SC 24, a 28% drop [27]. This means even lower energy CMEs in SC 24 are able to attain halo status within LASCO FOV. In addition, the leading-edge height at the time a CME attains halo status was found to be lower in SC 24 [27]. While the reduced speed of SC 24 halo CMEs is in line with that in the general population (see Fig. 4), the higher halo CME abundance is clearly related to the backreaction of the weakened heliosphere on CMEs. It is well known that the fraction of halo CMEs in a given population is a measure of how energetic those CMEs are [5]. The overabundance in SC 24 implies that the halo fraction should be higher for special populations of CMEs. For example, the fraction of halos in CMEs associated with cycle-23 DH type II bursts is 55%. In SC 24, the fraction increases to 63% [32].

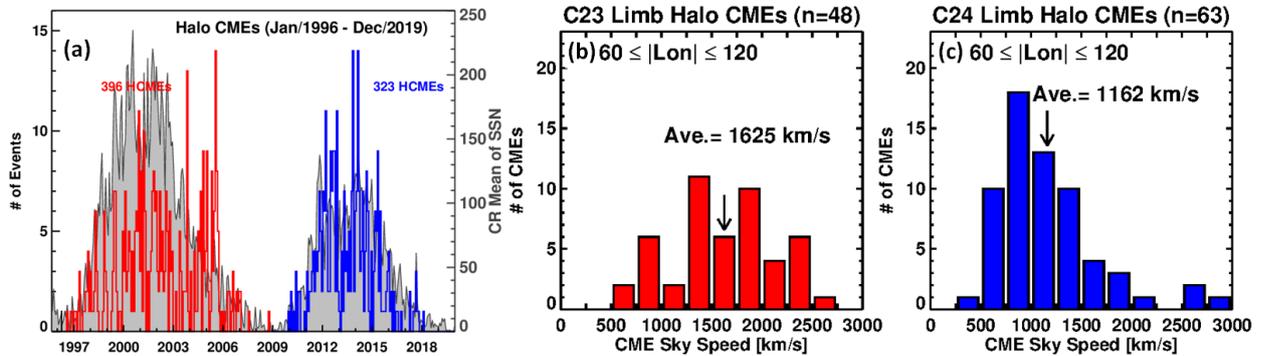

Figure 6. (a) Number of full halo CMEs observed by SOHO/LASCO summed over Carrington rotation periods in cycles 23 (red) and 24 (blue). (b) Speed distribution of limb halo CMEs in SC 23 and (c) in SC 24. The cycle-24 limb halo CME count declined by 21%; the CMEs are ~28% slower.

*4.2. Fast and Wide CMEs*

We see from 7a that the number of FW CMEs decreased from 485 in SC 23 to 253 in SC 24, by ~50%. Assuming that FW CMEs occurred at the same average monthly rate throughout SC 23 including the 4-month data gap in 1998 and 1999, then we need to add 13 FW CMEs, so the corrected number in SC 23 is 498. After this correction, the decrease of the number of FW CMEs in SC 24 is by 49%. The definition of FW CMEs is based on an early investigation DH type II bursts [33], which were associated with fast (speed ≥900 km/s and width > 60º). The drop is slightly deeper than that in SSN, most likely due to the fact that FW CMEs can also originate from non-spot regions. Furthermore, FW CMEs can originate even from behind the limb, so the correspondence with SSN variation need not be the same. FW CMEs play a central role in space weather events because their ability to drives shocks (and accelerate SEPs) and impact the magnetosphere causing geomagnetic storms.

*4.3. Reduction in SEP Activity and the Number of DH Type II Bursts*

During the first 4.5 years of SC 24, the decrease in number of large SEP events is only by 22% and the underlying CMEs are faster [6] when compared with the corresponding epoch in SC 23. When the whole cycles are compared, the number of large SEP events decreases from 102 in SC 23 to 46 in SC 24, amounting to a decrease by 55%, slightly more than that of FW CMEs (Fig. 7c). However, the reduction in the number of SEP events at higher energies (>500 MeV) is more drastic. As was noted in Fig. 2 and Table 1, there are only two GLE events in SC 24, compared to 16 in SC 23. The 88% reduction in the number of GLE events is a much greater drop than in FW CMEs or SSN.

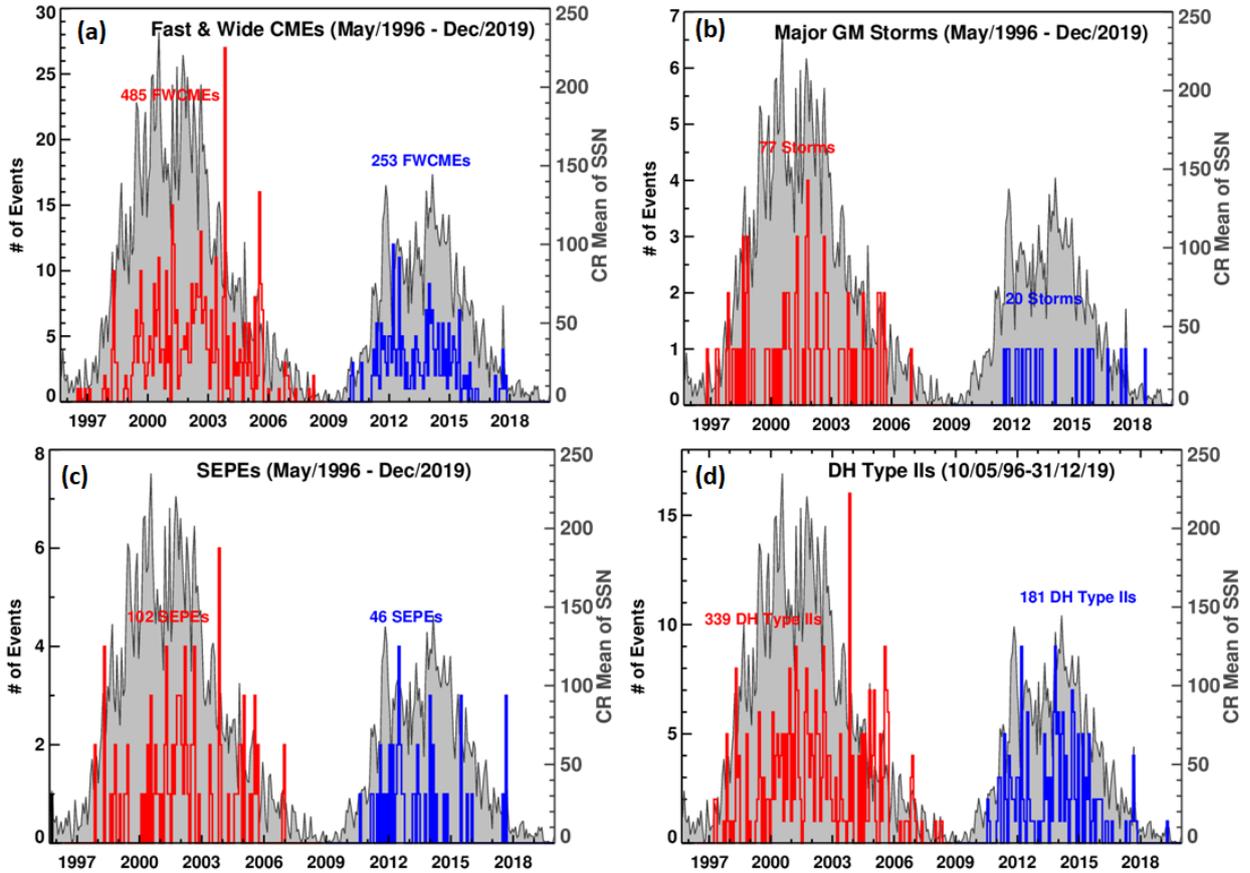

Figure. 7 (a) The rate of Fast and wide CMEs over Carrington rotation compared with that of (b) major geomagnetic storms (Dst < - 100 nT), (c) large SEP events (>10 MeV proton intensity ≥ 10 pfu), and (d) type II bursts observed in the decameter-hectometric (DH) wavelength range. These rates are superposed on the mean sunspot number over Carrington rotation(grey). In each case, SCs 23 (red) and 24 (blue) are distinguished and the total number of events in cycle are also noted.

Furthermore, the average speed of SEP-associated CMEs decreased in SC 24 in contrast to the first 4.5 years. Mewaldt et al. [34] reported on another measure of reduced SEP activity in SC 24 - the >10 MeV fluence, which drops by a factor of 5.8 as of October 2016 compared to that in SC 23 over the same epoch. In addition to the decrease in ambient magnetic field, these authors pointed to the lack of seed particles (due to reduced frequency of FW CMEs) as another factor that reduces the efficiency of SEP acceleration. Closely related to the SEP events are DH type II bursts caused by electrons accelerated in the same shock that accelerates SEPs. Figure 7d shows that the drop in the number of DH type II bursts is by 47% (from 339 in SC 23 vs. 181 in SC 24), which is very close to the drop in FW CMEs. This is understandable because the DH type II bursts are caused by FW CMEs [32-33]. Unlike SEP- and storm-producing CMEs, those causing DH type II bursts can occur anywhere on the disk (and even behind the limb), hence their close relation to FW CMEs.

*4.4. Interplanetary Shocks*

Shocks indicated by DH type II bursts and SEP events are observed in the solar wind by in-situ instruments as interplanetary (IP) shocks. Fast forward IP shocks detected by the Wind spacecraft have been compiled and made available at: https://lweb.cfa.harvard.edu/shocks/wi_data/. Most of them are driven by ICMEs, but we do not separate them here. There are 143 in SC 24 compared to 272 in SC 23, indicating a drop of 47%, which is very similar to that (48%) in the number of DH type II bursts or FW CMEs. In cycle 23, the number of Wind shocks in 2003 and 2004 seem to be too low, e.g., compared with the numbers from SOHO proton monitor data available from: http://umtof.umd.edu/pm/FIGS.HTML. If we correct for this difference, the drop in the number of SC 24 shocks is slightly higher. The speed distribution of the shocks differed significantly between the cycles, the mean values being 481 km/s (SC 23) and 374 km/s (SC 24). Thus, SC 24 witnessed fewer and lower speed shocks at 1 au, similar to the FW CMEs at the Sun. CMEs associated with limb flares of size ≥ C3.0 are useful in assessing the change in the CME speed near the Sun because the sky-plane speeds are closer to the true speeds [18]. Figure 8 shows the distributions of the speed and width of such limb CMEs: 601 in SC 23 and 407 in 24. There are ~32% lower number of such CMEs in SC 24, a less drop than the number of FW CMEs. The average speed drops significantly (by 18%) from 689 km/s to 568 km/s, somewhat similar to the drop (22%) in the IP shock speeds. On the other hand, the width increases by 14% because in SC 24, CMEs are wider due to the weak state of the heliosphere. The larger halo fraction in SC 24 (10% vs. 6% in SC 23) is also observed in other CME populations (e.g., those associated with IP type II bursts [32]) and is also attributed to the anomalous expansion of CMEs in SC 24.

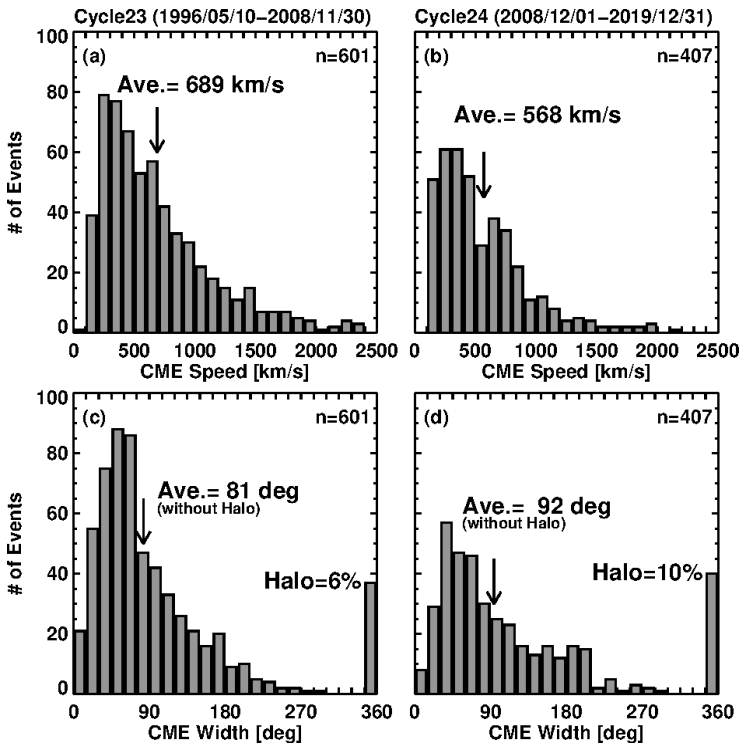

**Figure 8.** Distributions of the speed (a,b) and width (c,d) of limb CMEs in SCs 23 and 24. The total numbers are slightly different from [18] because of the cycle length used. The cycle averages (Ave.) of speeds and widths are noted on the plots. In the bottom plots, the width averages are shown excluding the full halo CMEs because their true widths are not known. SC 24 CMEs are slower but wider on the average.

Magnetosonic (fast) Mach numbers ($M_m$) are also available in the Wind shock data base noted above. The $M_m$ distributions are similar in the two cycles with nearly the same average values: 2.14 (SC 23) and 2.04 (SC 24). The constancy of average $M_m$ between the two cycles can be understood from the fact that drop in shock speed is balanced by the drop in magnetosonic speed (due to drop in temperature and magnetic field in the solar wind). At the Sun, where the solar wind is not fully developed, $M_m$ is determined primarily by the CME speed and the magnetosonic speed. Since the sound speed is typically ~100 km/s, the fast mode speed is close to the Alfven speed, $V_A$ (i.e., Alfvenic and fast Mach numbers are similar). If the drop in $V_A$ at 1 au (see Fig. 5) remains the same close to the Sun, it is countered by the drop in CME speed, so the fast Mach number remains roughly the same in the two cycles: $dM_m/M_m = dV/V - dV_A/V_A$. Recall that $dV_A/V_A = -0.15$ and $dV/V = -0.18$, so $dM_m/M_m = -0.026$.

*4.5. Energetic Storm Particle Events*

An important signature of IP shocks at 1 au is an energetic storm particle (ESP) event [35] consisting of particles accelerated in the shock locally. About 75% of IP shocks have energetic protons in the keV range, while the fraction drops to 45% in the MeV range [36]. A recent survey [37] identifies 95 ESP events accompanying SEP events during the years 1996-2017 using detection at two energy channels at 2 and 20 MeV. Over this period, about 400 IP shocks have been detected at L1. Thus, only about 24% IP shocks are associated with such SEP-ESP events. SEPs are accelerated near the Sun, where the shocks are strong, while the ESPs are accelerated by shocks at 1 au after they have evolved over a couple of days. ESPs are produced by shocks whose driving CMEs are more energetic as indicated by the CME speeds: 1088 km/s (ESP-producing CMEs) vs. 771 km/s (non-ESP CMEs). The difference is also indicated by the 1-au Alfvenic Mach numbers: 3.46 and 2.22 for ESP- and non-ESP shocks [36]. If we count only those ESP events that have an intensity $\geq$ 10 pfu in the >10 MeV GOES energy channel (see https://cdaw.gsfc.nasa.gov/CME_list/sepe/), we find 31 such ESP events, 22 in SC 23 and 9 in SC 24. All SC 24 events had associated SEP events, while 5 in SC 23 did not have an associated large SEP event (high background or the SEP event has intensity <10 pfu). Similar fractions of IP shocks produced SEP-ESP events in the two cycles: 9 out of 143 (or 6.3%) in SC 24 vs. 17 out of 272 or 6.3%). Only very energetic CMEs result in high-intensity ESP events as indicated by the CME speeds of these events: 1593 km/s (SC 24) and 1740 km/s (SC 23) (using the available speed measurements for 12 events in SC 23 and 9 events in SC 24). The drop in the number of high-intensity SEP-ESP events is 47%, which is smaller than the drop in the number of SEP events (see Table 1), but similar to that of FW CMEs, DH type II bursts, and IP shocks.

*4.6. Reduction in CME-related Geomagnetic Activity*

The reduction in the number of major geomagnetic storms (Dst $\leq$ - 100 nT) due to CMEs is by 74% (from 77 in SC 23 vs. 20 in SC 24, see Fig.7b). This is a more pronounced drop than the decrease in FW CMEs. While the reduction in intense storms is drastic, the drop in moderate geomagnetic storms in SC 24 is by 40%, similar to the drop in SSN [38]. These results have been recently confirmed [39-40] and other changes in geospace impact have also been reported. Kakad et al. [39] showed an increase in magnetopause standoff distance by ~4%, a significant decrease in Joule-heating of the polar ionosphere, and a drop in the strength of the equatorial electrojet.

It must be noted that the decrease in the number of intense and moderate storms is more than the decrease in the number of interplanetary CMEs (ICMEs) as well as the MC subset. Table 1 shows that the number of ICMEs dropped by 32%, while the number of MCs dropped by only 25%. This is consistent with the higher abundance of halo CMEs in SC 24, which when Earth-directed, are observed as ICMEs at Sun-Earth L1. The lower geoeffectiveness reflects the fact that the magnetic content of these ICMEs is diluted in SC 24. Furthermore, the average speeds of ICMEs are lower in SC 24. The reduction in the product VBz results in weaker storms. Note that we have not considered other factors such as rotation of MC axis [41] that can reduce the MC's geoeffectiveness.

*4.7. Reduction in CIR-related Geomagnetic Activity*

Intense storms can also be produced by CIRs formed due to the collision of fast solar wind streams with a slower stream ahead. Table 1 shows that the number of intense storms decreased from 12 in SC 23 to just 3 in SC 24, amounting to a 75% drop, similar to the drop in the number of CME storms. Table 2 lists the intense CIR storms from cycles 23 and 24. The list includes three storms with Dst >– 100 nT. The Dst indices of these storms were < –100 nT in the provisional Dst data. The final values become slightly higher, but we kept them in Table 2. In SC 24, one storm is in the maximum phase between the two SSN peaks and the remaining two are in the declining phase. Three of the 12 storms in SC 23, are in the rise phase and the remaining ones are from the declining phase. We have also listed the properties of the coronal holes (coronal hole area in EUV images and average field strength at the photospheric level). The coronal hole area is taken as the area in which the EUV intensity is less than a threshold value set as half the disk intensity in 284 Å image (SOHO/EIT in SC 23) and one third of disk intensity at 211 Å (SDO/AIA in SC 24). The 1998 August 7 storm has no measurements because it occurred during SOHO data gap. Table 2 shows that average area in the two cycles is roughly the same. However, the average magnetic field strength in SC 24 is 8.2 G compared to 13.7 G in SC 23. The decrease of 40% is almost the same as the drop in SSN. The average unsigned flux in SC 24 is $6.0\times10^{21}$ Mx compared to $1.76\times10^{22}$ Mx in SC 23, which amounts to a much larger drop by 66%. The number of events in Table 2 is too small for deriving statistical parameters. One way forward is to include more events by extending the analysis to moderate storms.

Table 2. List of intense storms in cycle 23 and 24 and the associated coronal hole properties

| Dst Date | Time | Dst (nT) | CH Date & Time | Loc. | Pol. | Area ($10^{20}$cm$^2$) | $<|B|>$ (G) | Flux ($10^{21}$ Mx) |
|---|---|---|---|---|---|---|---|---|
| Cycle 23 | | | | | | | | |
| 1996/10/23 | 05:00 | -105 | 1996/10/20 07:03 | N00 | + | 7.33 | 7.3 | 5.35 |
| 1998/03/10 | 21:00 | -116 | 1998/03/08 09:03 | S30 | - | 1.86 | 11.9 | 2.22 |
| 1998/08/07 | 06:00 | -108 | 1998/08/04 04:14 | ??? | + | ---- | ---- | ---- |
| 2002/09/04 | 06:00 | -109 | 2002/08/31 06:48 | S15 | + | 9.39 | 15.6 | 14.7 |
| 2002/10/07 | 08:00 | -115 | 2002/10/05 01:48 | S07 | + | 24.6 | 20.0 | 49.2 |
| 2002/10/14 | 14:00 | -100 | 2002/10/11 02:39 | N25 | - | 8.50 | 18.9 | 16.1 |
| 2002/11/21 | 11:00 | -128 | 2002/11/18 13:21 | S04 | + | 8.40 | 12.5 | 10.5 |
| 2003/07/12 | 06:00 | -105 | 2003/07/07 21:40 | N04 | - | 6.61 | 10.9 | 7.20 |
| 2004/02/11 | 18:00 | -93 | 2004/02/10 09:24 | N02 | - | 17.6 | 11.2 | 19.7 |
| 2005/05/08 | 19:00 | -110 | 2005/05/07 01:36 | N10 | - | 20.6 | 16.5 | 34.0 |
| 2005/08/31 | 20:00 | -122 | 2005/08/29 10:48 | S12 | + | 18.0 | 17.0 | 30.6 |
| 2006/04/14 | 10:00 | -98 | 2006/04/12 17:59 | N02 | - | 4.77 | 9.2 | 4.41 |
| Cycle 24 | | | | | | | | |
| 2013/06/01 | 09:00 | -124 | 2013/05/30 00:00 | N00 | + | 10.3 | 9.3 | 9.62 |
| 2015/10/07 | 23:00 | -124 | 2015/10/05 08:00 | S05 | + | 9.60 | 7.3 | 7.03 |
| 2016/03/06 | 22:00 | -98 | 2016/03/02 17:15 | S08 | - | 17.0 | 8.0 | 1.36 |

**5. Expected Space Weather in Cycle 25**
Given the mild space weather in SC 24, what is the prognosis for SC 25? There have been a number of attempts to predict the strength of SC 25 (see e.g., [42-43] and references therein). One such attempt is to use Sun's polar microwave brightness at 17 GHz, which is a direct indicator of the polar magnetic field strength [44-45]. The method is illustrated in Fig. 9. In Fig. 9a we have plotted the 17 GHz brightness temperature (Tb) values averaged over latitudes >60⁰. Figure 9a shows the smoothed Tb as a function of time with the peak Tb values in the three minima (22/23, 23/24, 24/25) with the months of occurrence of the peaks marked. The 13-month smoothed total sunspot number (TSSN) from SILSO is shown for comparison. The low-latitude (LL) Tb, which is a proxy to the hemispheric sunspot number (HSSN) is also shown in Fig. 9a. The LL Tb peaks at different times in the two hemispheres and is dominant in the southern hemisphere both in cycles 23 and 24. Furthermore, the second peak in TSSN roughly coincides with the LL Tb peak in the southern hemisphere. The HL Tb peaks at least a couple of years before the SSN minima and that the peak is rather flat after the initial steep rise. This provides an opportunity for estimating the next cycle strength well ahead of time. Based on the correlation between polar Tb and the polar magnetic field strength [44], and the result that the peak Tb during a given solar minimum is correlated with the peak HSSN in the subsequent solar maximum give a pathway to predict the HSSN of a cycle from the Tb peak of the previous minimum. Table 3 lists 6 peak Tb values for the three solar minima: 22/23, 23/24, and 24/25 in both hemispheres. Also listed are the 4 HSSNs from SILSO for cycles 23 and 24. Figure 9b shows the scatter plot between HSSN and peak Tb (in $10^4$ K) with regression line,

$$HSSN = 931.6 \times Tb - 982.9. \qquad (1)$$

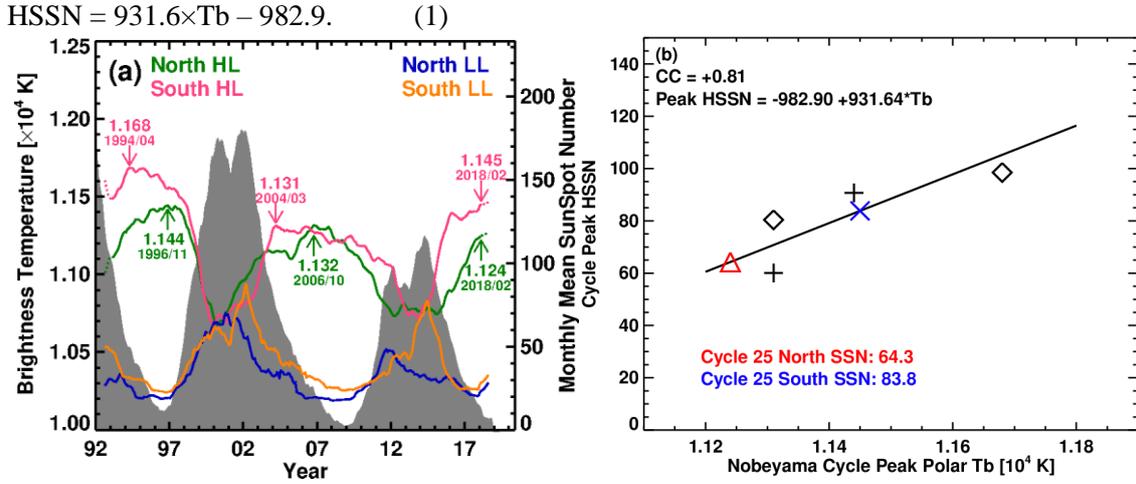

Figure 9. (a) high (HL) and low latitude (LL) brightness 17 GHz brightness temperature (Tb) from the Nobeyama Radioheliograph (NoRH) plotted as a function of time for northern and southern hemisphere of the Sun. Monthly Tb values from latitudes >60⁰ are smooth it over 13 months. Cycle-peak values of HL Tb in cycles 22, 23, and 24 are noted on the plots along with the year and month. (b) Treating each hemisphere independently, the relation between the peak Tb in a cycle with the peak hemispheric sunspot number (HSSN) of the following cycle is derived using data from cycles 22 and 23. The diamonds and + symbols denote data points from the northern and southern hemispheres, respectively. The red triangle and blue cross mark the predicted cycle-25 hemispheric SSN in the northern and southern hemisphere, respectively.

| Table 3. Peak Tb values for the three minima and the hemispheric SSN | | | | | | |
|---|---|---|---|---|---|---|
| | Cycle 22/23 | | Cycle 23/24 | | Cycle 24/25 | |
| Hemisphere | North | South | North | South | North | South |
| Peak Tb [$10^4$ K] | 1.17 | 1.14 | 1.13 | 1.13 | 1.12 | 1.15 |
| Tb peak month | 1994/04 | 1996/11 | 2004/03 | 2006/10 | 2018/02 | 2018/02 |
| Peak HSSN | 90.70 | 98.50 | 60.10 | 80.40 | | |
| Peak HSSN month | 2000/12 | 2002/03 | 2011/08 | 2014/04 | | |
| Predicted HSSN | | | | | 64.3 | 83.8 |

Now we can use the two peak Tb values corresponding to the 24/25 minimum to predict the cycle-25 HSSN as 64.3 (north) and 83.8 (south). Simply adding the two HSSN gives the summed SSN (SSSN) as an estimate of the TSSN, viz., 148.1. Another way is to statistically estimate TSSN from the observed TSSN and HSSN during past cycles. For this purpose, we make use of the HSSN and TSSN data available from three observatories: National Astronomy Observatory of Japan (NAOJ), Kanzelhöhe Solar Observatory (KSO), and SILSO. The NAOJ SSN data are monthly averages (see https://solarwww.mtk.nao.ac.jp/mitaka_solar1/data03/sunspots/number/Readme.txt) derived from sunspot drawings and are available since March 1939 until June 1998, and CCD-based digital data after June 1998 [46]. We smooth these data over 13 months. The KSO hemispheric data have been used before for studies of north-south asymmetry in SSN [47-48]. KSO data are available since 1944 with some data gaps in SC 20. The SILSO hemispheric sunspot data are available only from June 1992, while TSSN data are available over a longer period (http://www.sidc.be/silso/datafiles). We use the 13-month smoothed HSSN and TSSN made available on the SILSO web site. Figure 10a,b shows the NAOJ, KSO and SILSO time series data. There are clear differences among the different time series (probably due to differing methods of counting the solar features), but the hemispheric numbers are related to the total numbers in a similar way. Combining all three data sets, we get,

TSSN = 2.09 +0.91SSSN.  (2)

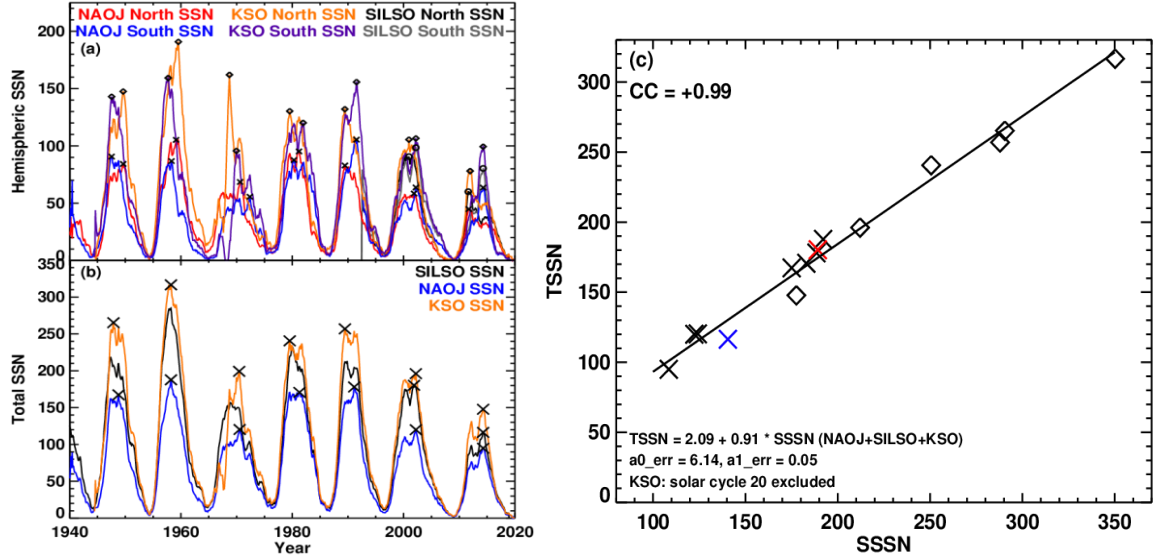

Figure 10. (a) HSSN from NAOJ, KSO, and SILSO. The HSSN peaks in each cycle are identified and added together to get the summed HSSN (SSSN). (b) TSSN from the three observatories with the peak values marked by crosses. (c) scatter plot between TSSN and SSSN using the combined data set: NAOJ (black crosses), KSO (diamonds), and SILSO (red and blue crosses denoting SC 23 and 24, respectively). The regression line is given on the plot. The correlation is high with $p < 0.005$. The error in the intercept (a0_err) and slope of the regression line (a1_err) are used to estimate the uncertainty in the prediction.

We use equation (2) to predict TSSN in SC 25 from the two HSSN values obtained from equation (1) using Tb from the SC 24/25 minimum. From Table 3, SSSN = 64.3 + 83.8 = 148.1, which when substituted in eq. (2) gives TSSN = 137. The uncertainty range obtained from the errors in the fit coefficients is from -14 to +13. The predicted TSSN for SC 25 is only slightly higher than the observed cycle-24 TSSN (116), but much smaller than the observed cycle-23 TSSN (180). A detailed discussion of various methods of predicting cycle strengths can be found in [43]. Of these, physics-based methods indicate that the strength of SC 25 strength may not be too different from that of SC 24. We used the polar microwave brightness temperature to infer the polar magnetic field strength using which we predicted SC 25. This method can be used to predict HSSN when accurate polar field measurements are possible. As for space weather, one expects it to be milder in SC 25 as well because of the expected weak heliospheric total pressure, weak magnetic field strength in CMEs, sheaths, CIRs, and the ambient medium.

## 6. Summary and Conclusions

In this article, addressed the connection between solar activity and space weather events in SCs 23 and 24. Although the overall behavior of space weather events is similar to the solar activity represented by SSN, there are differences when we look closely at the individual activity peaks within a cycle. SEP events are generally concentrated around such peaks, which may be primary, secondary, or other peaks in SSN. The decline of solar activity in SC 24 seems to have two effects (i) reduced number of energetic solar events such as flares and coronal mass ejections, and (ii) the backreaction of the heliosphere on CMEs. The combined effect is to reduce high-energy space weather events more than the decline in SSN: the number of high-energy SEP events (e.g., GLE events) and intense ($Dst \leq -100$ nT) geomagnetic storms (Dst index is an indicator of ring current energy). Other phenomena related to CME-driven shocks such as interplanetary type II bursts, IP shocks, and high-intensity ESP events faithfully follow FW CMEs. The number of large SEP events declined slightly more than that of FW CMEs probably because of connectivity issues. The numbers of halo CMEs and interplanetary CMEs decline less severely than SSN because the weak heliospheric state causes CMEs to appear larger in SC 24. Some quantities like the magnetosonic Mach number remains roughly constant in the two cycles due to a balance between the decline in shock speed and the magnetosonic speed. Examining the properties of coronal holes responsible for intense CIR storms, we find that the underlying photospheric magnetic field is weaker in SC 24 by the same amount as the SSN. Using the polar microwave emission method, which predicts hemispheric SSN, we obtain the strength of SC 25 as ~137, which is only slightly larger than that of SC 24. This suggests that the space weather in SC 25 is expected be milder as well.

**Acknowledgments**

We benefited from the open data policy of SOHO, Wind, STEREO, SDO, and Wind missions. This work was supported by NASA's Living With a Star program. We acknowledge the following data sources. https://lasco-www.nrl.navy.mil/solwind_transient.list (Solwind CMEs1979 - 1985); https://www2.hao.ucar.edu/mlso/solar-maximum-mission/smm-cme-catalog (SMM C/P CMEs 1985 - 1989); https://cdaw.gsfc.nasa.gov/CME_list (SOHO/LASCO CMEs 1996 - 2019). https://cdaw.gsfc.nasa.gov/CME_list/halo/halo.html (Halo CMEs); https://cdaw.gsfc.nasa.gov/CME_list/sepe/ (SEPs); https://wwwbis.sidc.be/silso/infosnmtot (V2.0 SSN); http://wdc.kugi.kyoto-u.ac.jp/dstdir/ (Dst).